# The synergy of electromagnetic effects and thermophysical properties of metals in the formation of laser induced periodic surface structures

GEORGE, D TSIBIDIS[1,2a], PANAGIOTIS LINGOS[1b], AND EMMANUEL STRATAKIS[1,3c]

[1]*Institute of Electronic Structure and Laser (IESL), Foundation for Research and Technology (FORTH), Vassilika Vouton, 70013, Heraklion, Crete, Greece*
[2]*Department of Materials Science and Technology, University of Crete, 71003, Heraklion, Greece*
[3]*Department of Physics, University of Crete, 71003, Heraklion, Greece*

[a] *e-mail: tsibidis@iesl.forth.gr;*  [b] *e-mail: plingos@physics.uoc.gr;*  [c] *e-mail: stratak@iesl.forth.gr*

**Femtosecond pulsed lasers have been widely used over the past decades for precise materials structuring at the micro- and nano- scales. In order, though, to realize efficient material processing and account for the formation of laser induced periodic surfaces structures (LIPSS), it is very important to understand the fundamental laser-matter interaction processes. A significant contribution to the LIPSS profile appears to originate from the electromagnetic fingerprint of the laser source. In this work, we follow a systematic approach to predict the *pulse-by-pulse* formation of LIPSS on metals due to the development of a spatially periodic energy deposition that results from the interference of electromagnetic far fields on a non-flat surface profile. On the other hand, we demonstrate that the induced electromagnetic effects, alone, are not sufficient to allow the LIPSS formation, therefore, we emphasize on the crucial role of electron diffusion and electron-phonon coupling on the formation of stable periodic structures. Gold and stainless Steel are considered as two materials to test the theoretical model while simulation results appear to confirm the experimental results that, unlike gold, fabrication of pronounced LIPSS on stainless Steel is feasible.**

The employment of femtosecond (fs) pulsed laser sources for material processing has received significant attention due to the important technological applications [1-5]. Various types of surface nano/micro-structured topographies have been fabricated by exploiting the wealth of possibilities laser technology offers through modulation of the laser parameters (such as the energy, photon energies, polarization states, energy dose, sequence of pulses, etc.). As a result, a plethora of periodic or aperiodic patterns have been produced that are biomimetic (i.e. they resemble the topographies and exhibit impressive functionalities of the surface of various species found in nature) [6]. Due to the unequivocal precision of the technique and the evident capabilities, it is of paramount importance to control the laser energy and characteristics in a way that will allow fabrication of such patterns in a systematic fashion. This endeavor requires a deep understanding of how laser irradiation interacts with matter.

Despite the presence of several theoretical frameworks that aim to elucidate the surface modification related underlying physical processes [7,8], it has been postulated that the pattern formation is attributed to a great extent to the electromagnetic fingerprint of the solid. In principle, the formation of various types of periodic patterns ranging from Low (LSFL) [9,10] to High Spatial Frequency Laser Periodic structures (HSFL) [1,8,11] or even more complex structures [12] have been predominantly attributed to the excitation of scattered surface waves, including Surface plasmons [9,13] while the development and coupling of near/far fields with the incident beam leaves an electromagnetic fingerprint on the solid [14]. However, experimental results for various metals showed that while the size and origin of the periodicity of LIPSS can be ascribed to electromagnetic effects, the manifestation of the depth profile of the produced periodic structures varies [15]. Although, one might argue that, the electromagnetic response of the irradiated material (i.e. different extinction coefficient) accounts for this behavior, experimental observations demonstrate that other physical mechanisms can be potentially be involved that allow or inhibit the formation of distinct periodic structures [15,16]. One possible candidate to explain the experimental results is related to the propagation of the laser energy inside the material and associated thermal effects.

In this Letter, we demonstrate in a conclusive way, that despite the importance of the electromagnetic effects that take place after irradiation of conducting surfaces (i.e. metals) on the prescribed features of the surfaces waves, the thermophysical properties of the material themselves play a crucial role in the formation of periodic patterns. To this end, we have conducted a multiscale physical modelling approach to describe pattern formation on two materials, gold (Au) and stainless Steel (SS) that are characterized from distinctly different thermophysical properties. Experimental observations in previous reports [16] have confirmed that in bulk solids, while pronounced LSFL are formed on SS [17], a very shallow rippled topography is produced in Au [16].

Our theoretical predictions are aimed to reveal the significant role of the differences in the electron diffusion and electro-phonon coupling strengths in the two materials to the features of the induced topographies. Although some aspects of the elucidation of the impact of the electron diffusion and electron-phonon coupling have been addressed in a previous report [15], in this Letter, we want to emphasize the interplay between the electromagnetic and thermophysical properties in the produced amplitude of periodic structures through a detailed theoretical investigation. In our simulations, linearly polarized laser beams of pulse duration 170 fs and wavelength $\lambda_L$=513 nm were used. It is noted that the electric field is polarized along the *X*-axis.

To describe, firstly, the electromagnetic response of the two materials, lightly rough surfaces are assumed (of roughness size much smaller than the laser wavelength size). The aim is to show that non-flat topographies account for the excitation of electromagnetic modes that are precursors of periodic structures. To emulate the rough patterns, which are produced upon irradiation within the first number of pulses, a configuration of a random distribution of sub-wavelength hemispherical holes (radius randomly ranged between $10 < R < 100$ nm) is used along the surface. A similar approach has been introduced in previous reports [14].

The interaction of light with the surface inhomogeneities of the material produces electromagnetic interference patterns along the surface which determines the energy absorption landscape. Hence, the investigation of the periodic (or quasi-periodic) electromagnetic modes along the surface is the first step in our multi-physical study. The interference between the incident light beam with the scattered waves generated by surface nano-defects on metallic surfaces as a result of dipole-dipole coupling, produce standing mixed waves such as Surface Plasmon Polaritons (SPP) and quasi-cylindrical evanescent waves [18]. The optical properties of the irradiated material determine the characteristics of these waves such as scattering, absorption, transmission, the optical propagation length as well as the skin depth. To reveal the electromagnetic response of the material, a Finite Integration Technique is employed for the solution of the full-vector 3D Maxwell-Grid Equations (see Suppementary Matwrial) [19-22]. Assuming a laser beam at a fixed wavelength $\lambda_L = 513$ nm, the dielectric constants considered in the simulations are $\varepsilon_{Au} = -3.43 + i3.01$ and $\varepsilon_{SS} = -0.45 + i16.3$ for Au and SS [23,24], respectively, whereas the hemispherical nano-holes are assumed to be filled with air ($\varepsilon = 1$). The electric filed of the laser beam is presented as a plane-wave linearly polarized along the *X*-axis arriving at normal incidence to the surface. The geometry of the problem allows to use a simple representation of the boundary conditions and the computational grid is terminated by two convolutional perfectly matched layers (convPML) [25] along the *Z*-direction to avoid non-physical reflections while periodic boundary conditions are used for *X* and *Y* directions.

It is noted that in order to compare the surface deformation of both materials following the multi-pulse irradiation, we used identical surface topographies before the arrival of the initial laser pulse as well as the same irradiation conditions. To acquire information about the energy absorption in these two samples, we illustrated the inhomogeneous distribution of the intensity, $I \sim |\vec{E}|^2$ (where $\vec{E}$ stands for the electric field), below the rough surface normalized with respect to the maximum intensity of the flat surface $I_S$ or the maximum intensity of the incident beam by the source. In Fig.1a-bb, we show the normalized intensity difference $(I - I_S)/I_S$ below the surface in the transverse plane perpendicular to light wave-vector, which depicts the intensity maxima and minima due to both scattered radiative and non-radiative fields by the subwavelength imperfections. It appears, that the presence of the nano-holes in the metallic surfaces favor the energy deposition at their edges, localized parallel to the laser polarization, while the far fields consisting of SPP and quasi-cylindrical wave components exhibit maxima and minima that are formed perpendicularly to the laser polarization. Previous simulations confirmed that near field enhancement effects on such nano-holes contribute to the formation of sub-wavelength HSFL via convection flow process [11]. For both metals in the current work, the superposition of the incident beam with the scattered far fields determines the energy absorption maps along the surface. It is noted that there exists a strong difference in the optical properties and response of the materials: (a) the absorption coefficient (b) the SPP propagation length and (c) the skin depth. For instance, the absorption coefficient of is $a_{Au} = 81.73\ \mu m^{-1}$ and $a_{SS} = 47.75\ \mu m^{-1}$ for the two materials. The SPP propagation length is $L_{Au} \approx 100\ \mu m$ for Au and $L_{SS} = 0.23\ \mu m$ for SS. Despite these differences, the absorption features of these materials are quite similar determined by the radiative far fields. On the other hand, the far field couplings are negligible because the average distance between the nano-holes is nearly equal to the laser wavelength. As a result, the periodic/quasi-periodic absorption features are observed with average periodicity $\approx \lambda_L$. This is also captured in Fast Fourier Transforms of the corresponding absorption maps ($k_x/k_0 \approx \pm 1$) (Fig.1c-d). Moreover, the periodic absorption features are strongly defined by the SPP. In Fig.1e-f, we present the normalized *Z*-component of the electric field at the propagation plane *XZ* which depicts the SPP along the air-metal interface; in the case of Au, the SPP is pronounced compared to the stainless steel. According to the above discussion, the presence of SPP is expected to play a dominant role on the pulse-by-pulse modification of the material's surface since it dictates the thermo-affected/melt regions which evolve the topography of the surface.

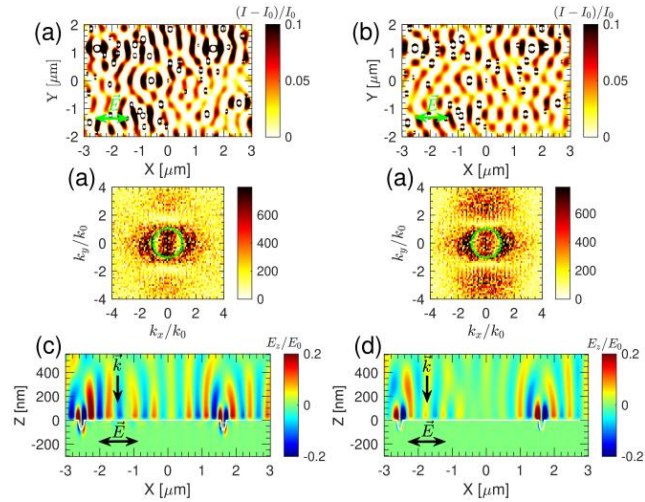

**Fig.1.** Absorbed energy distributions on the transverse plane for Au (a) and SS (b) surfaces. (c), (d) illustrate the Fourier transform of (a), (b), respectively. The *green* circles represent the boundary of $|k| = k_0$ where $k_0 = 2\pi/\lambda_L$ stands for the wave-vector of light propagation in air (double headed arrow indicates laser polarization direction). (e) and (f) illustrate normalized *Z*-component of the electric field at the propagation plane *XZ*. (double-headed arrows indicates laser polarization direction). In this cross section, two nano-holes are located at positions *X*=-2540 nm and *X*=1630 nm. The *white* line represents the air-metal interface.

To correlate the electromagnetic response of the material with changes occurring along the irradiated surface in a pulse after pulse approach, a side view of the distribution of the absorbed energy is illustrated for both materials on the *XZ* plane (the propagation plane) for four number of pulses (*NP*) (Fig.2a and Fig.2b). The periodic modulation of the absorbed energy due to the interference effects shown in Fig.1 is also depicted on the propagation plane. The electric field (*Z*-component) distribution for the two materials is also illustrated in Fig. 2c and Fig.2d. Results for Au and SS demonstrate similar far field interference patterns of the absorbed energy, following scattering of the electromagnetic fields off the hole scatterers. Therefore, a reasonable outcome of such electromagnetic field distribution would be the formation of periodic structures (LSFL) on the surface of both materials. Nevertheless, experimental results indicate that, unlike in SS, the amplitude of the ripples formed on bulk Au is very small even at very large number of pulses [26]. This outcome suggests that other effects related to the properties of the metals and thermal response of system after absorbing the laser energy might account for the distinct differences.

To describe the thermal response of the system and the process towards pattern formation, electron excitation, electron-phonon coupling, thermal effects and phase transitions are incorporated into a multiscale theoretical framework; the theoretical model comprises a Two Temperature Model (TTM) coupled with a Navier-Stokes equation that provides a detailed analysis of the fluid dynamics [9,11]. The source term entering the TTM which relates with the laser energy that excites the electron system is derived from the spatial intensity distribution of the electric field while a temporal dimension is also provided by scaling the intensity with $e^{-(t-3\tau_P)^2/\tau_p^2}$. Due to the fact that surface modification is necessary to induce pattern formation, the fluence of the laser beam is adjusted appropriately to lead to a phase transition. The phase transition is based on a thermal criterion that the lattice temperature exceeds the melting point of the material. Appropriate laser conditions are considered in this study to ensure that a substantial portion of fluid volume is produced, which could eventually lead to pattern formation, via fluid transport. . A systematic study of the ultrafast phenomena and relaxation processes for the two metals indicate the following important results: due to the large electron heat diffusion and ballistic transport for Au, less energetic electrons remain on the surface of the material to couple with the lattice system. By contrast, the behavior of SS, both in terms of electron diffusion and electron-phonon coupling strength, is different: the electron heat conductivity length is four times smaller than those for Au while the electron phonon coupling strength is ten time stronger [17] and it follows a decreasing monotonicity with increasing electron temperature (opposite to that of Au [27]). As a result, larger lattice temperatures develop in a smaller region, thus the developed temperature gradients are more enhanced and finally, the induced hydrothermal waves will have larger amplitudes, giving rise to pronounced periodic profiles upon resolidification (see Supplementary Material).

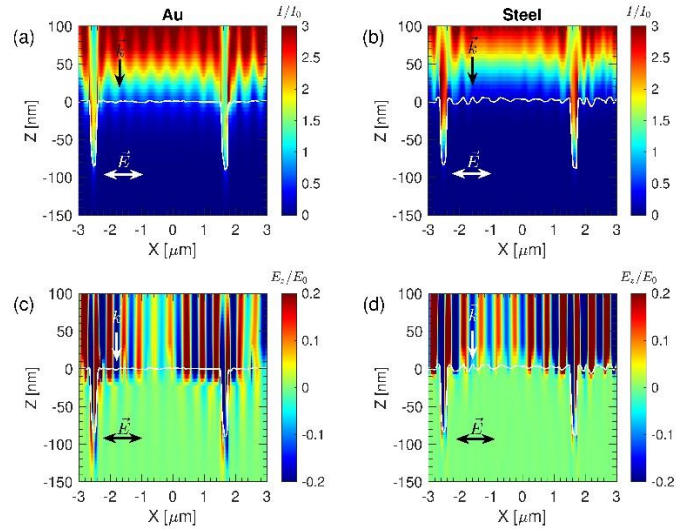

**Fig.2.** Absorbed energy distributions on the propagation plane for (a) Au and (b) SS surfaces (*NP*=4). Electric field (*Z*-component) distribution for Au (c), SS (d). The *white* line defines the boundary of the surface profile along the air-metal interface.

Given that the electrons lose energy through two competing processes, the electron diffusion and electron-phonon coupling, the presence of low energetic electrons (in a volume near the surface) that will interact through scattering with the phonon system and the weak electron-phonon coupling for Au [27] will lead to small lattice temperatures and thermal gradients close to the surface; to demonstrate the lattice energy distribution inside the volume of the irradiated materials, 1D simulations have been conducted to give a spatiotemporal profile of the lattice temperature (Fig. 3). On the other hand, the (spatially) periodic energy modulation resulting from the interference effects will be transferred to the electron system which implies the electron temperature will exhibit a periodic distribution (see Supplementary Material). The lattice system will acquire a similarly periodic spatial modulation through electron-phonon scattering and the strength of the scattering and the lattice temperature values will be the critical factors that will determine the fluid movement and whether the induced hydrothermal waves will result into stable and visible rippled structures.

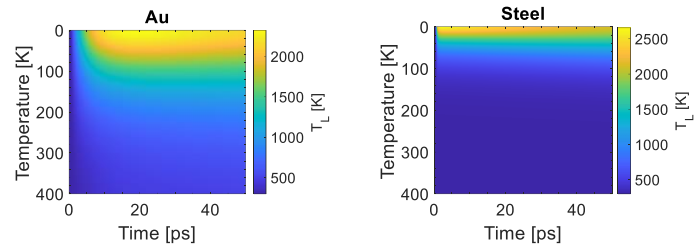

**Fig.3.** Spatiotemporal profile of lattice temperature for Au (a) and SS (b).

To illustrate the effect of repetitive irradiation that will be projected on the surface topography, we have carried out simulations at increasing *NP*. Results show an enhanced absorption in the wells of the periodic profile (Fig.2) and the excitation of surface plasmon waves is followed from a periodic energy absorption. As repetitive irradiation is expected to increase the amplitude of the induced periodic structures due to mass transport, a repetitive modelling scheme was applied to predict the variation of the height and periodicity of the ripples (Fig.4). Simulations results for both Au and SS show a shallow rippleprofile for Au (Fig.4a), compared to higher ripple amplitudes for SS (Fig.4b) due to the predominantly energy confinement for SS as explained above. Although simulations shown in Fig.4 are results for *NP*=4, similar conclusions can be deduced at higher *NP*. With respect to the periodicity variation at increasing *NP*, it is noted that results in previous works showed a decrease of the ripple periodicity as the energy dose *NP* increases; this behavior results from a shift of the wavelength of the excited SPP to smaller values when the profile becomes deeper at increasing *NP* [18,28,29]; by contrast, in the aforementioned investigation, our simulations indicate that for a small number of pulses, an apparent ripple periodicity reduction is not expected (Fig.4c). Given that the emergence of a pronounced decreasing monotonicity appears following irradiation of more rough profiles (or regions that have undergone ablation) [18,28,29], a future investigation might involve the application of the model in extreme laser conditions that involve mass removal. Finally, the substantial increase of the amplitude of the produced ripples for SS, in contrast to Au, at increasing *NP* is illustrated in Fig.4d.

One important outcome from the above discussion is that nano/micro-patterning of materials could be tailored even though the thermophysical properties of the solid are not sufficiently strong to lead to remarkably pronounced topographies (in the case of Au or other noble metals exhibiting small ripple amplitude). More specifically, given the significant impact of the thermophysical properties of a material in the produced topography, electron diffusion length can be modulated by inhibiting, for example, the electron transport inside the volume (i.e. nano-structuring of thin films) [30,31]. Similarly, the relaxation process of the electron system into an equilibrium and the acceleration of the scattering effects can also be controlled through selecting appropriate metal thicknesses [31]. This will lead to an appropriately tailored exchange of energy between the electron and lattice systems when the periodic modulation of the electron temperature is high enough; as a result, the produced hydrothermal waves are expected to lead upon resolodification to rippled patterns of enhanced amplitude sizes.

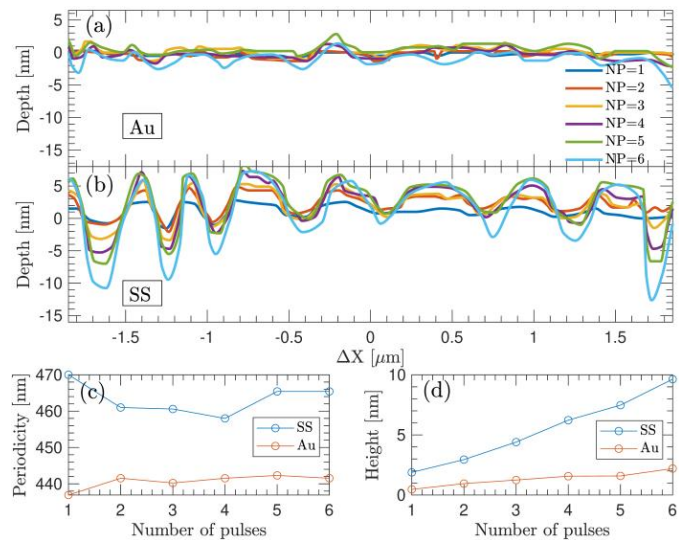

**Fig.4.** Amplitude of LIPSS as a function of number of pulses for (a) Au and (b) Stainless steel. Average periodicity (c) and (d) amplitude of LIPSS *vs* number of pulses.

In conclusion, our approach aimed to interpret the apparent differences in the periodic pattern amplitudes following irradiation of materials with distinctly different thermophysical properties. The study: (i) predicts the excitation of electromagnetic modes of periodicities of the size of the wavelength (consistent with the periodicities of LSFL structures), (ii) demonstrates the significant influence of the electron diffusion and electron-phonon coupling on the quantitative features of the expected periodic patterns and (iii) determines whether the irradiated material will sustain the formation of periodic structures. Following the major impact of the material properties in nanostructuring, the ability to control the thermal response of a material can provide novel routes for optimizing the outcome of nano-processing and fabricating application based biomimetic structures.

**Funding**. *BioCombs4Nanofibres* (grant agreement No. 862016); *NEP* project (GA 101007417); COST Action *TUMIEE*.

**Disclosures.** The authors declare no conflicts of interest.

**Data availability.** Data underlying the results presented in this paper are not publicly available at this time but may be obtained from the authors upon reasonable request.

*Supplementary Material*

*to*

# The synergy of electromagnetic effects and thermophysical properties of metals in the formation of laser induced periodic surface structures

GEORGE , D TSIBIDIS[1,2] , PANAGIOTIS LINGOS[1] , AND EMMANUEL STRATAKIS[1,3]

[1]*Institute of Electronic Structure and Laser (IESL), Foundation for Research and Technology (FORTH), Vassilika Vouton, 70013, Heraklion, Crete, Greece*
[2]*Department of Materials Science and Technology, University of Crete, 71003, Heraklion, Greece*
[3]*Department of Physics, University of Crete, 71003, Heraklion, Greece*

## I. Numerical scheme of the electromagnetic effects

A precise evaluation of the electromagnetic fingerprint of laser irradiation of a solid requires the solution of Maxwell equations on the affected region.  An accurate description of the induced electromagnetic modes that are developed is expected to lead to a precise computation of the spatial distribution of the absorbed energy on the irradiated solid. To calculate, precisely, the electromagnetic fields distribution along the whole material, a computational approach based on the Finite Integration Technique (FIT) is employed [1-3]. This method shares similarities in accuracy with the widely used

Finite Difference Time Domain method (FDTD) [4] for the solution of the Maxwell's equations when it is implemented on Cartesian coordinate grids. FIT is based on solving the Maxwell's equations in their integral form:

$$\oint_{\partial A} \mathbf{E} \cdot d\mathbf{s} = -\iint_A \frac{\partial \mathbf{B}}{\partial t} \cdot d\mathbf{A} \quad (1.a)$$

$$\oint_V \mathbf{B} \cdot d\mathbf{a} = 0 \quad (1.b)$$

$$\oint_{\partial A} \mathbf{H} \cdot d\mathbf{s} = \iint_A \left(\frac{\partial \mathbf{D}}{\partial t} + \mathbf{J}\right) \cdot d\mathbf{A} \quad (1.c)$$

$$\oint_V \mathbf{D} \cdot d\mathbf{A} = \iiint_V \rho dV \quad (1.d)$$

where the field and flux vectors are related to the material equations:

$$\mathbf{B} = \mu \mathbf{H} + \mathbf{M}, \qquad \mathbf{D} = \varepsilon \mathbf{E} + \mathbf{P}, \qquad \mathbf{J} = \sigma \mathbf{E}, \quad (2)$$

FIT employs a pair of staggered grids, the primary grid $G$ and the dual grid $\tilde{G}$ such as the Yee cells in FDTD method where each edge of one grid penetrates a surface of the other grid. The primary grid represents the entire computational domain as a collection of volume cells $V_i (i = 1 \ldots N_V)$ surrounded by facets $A_i (i = 1 \ldots N_A)$ and edges $L_i (i = 1 \ldots N_L)$. In each cell the material is assumed homogenously distributed. In order to construct the numerical form of the Maxwell's equations, a type of state variables is needed to be defined. In FIT, these state variables are called *grid voltages* and *grid fluxes*, an integral form of the electric and magnetic field vectors along the grids:

$$\mathbf{e}_i = \int_{L_i} \mathbf{E} \cdot d\mathbf{s} \qquad \mathbf{h}_i = \int_{\tilde{L}_i} \mathbf{H} \cdot d\mathbf{s} \qquad \mathbf{d}_i = \int_{\tilde{A}_i} \mathbf{D} \cdot d\mathbf{A} \qquad \mathbf{j}_i = \int_{\tilde{A}_i} \mathbf{J} \cdot d\mathbf{A} \quad (3)$$

Here, for instance, the electric voltage $e$ related to one edge of the surface, represents the exact value of the integral of the electric field strength along that path, while $b$ represents the exact value of the magnetic flux density integral over a cell surface. Using the above state variables, the integral equations Eq.(1) can be transformed into the corresponding discrete grid form known as *Maxwell's Grid Equations*:

$$\mathbf{Ce} = -\frac{d}{dt}\mathbf{b} \quad (4.a)$$

$$\mathbf{Sb} = 0 \quad (4.b)$$

$$\mathbf{Th} = \frac{d}{dt}\mathbf{d} + \mathbf{j} \quad (4.c)$$

$$\tilde{\mathbf{S}}\mathbf{d} = q \quad (4.d)$$

where $\mathbf{C}$ is an $N_A \times N_E$ matrix which represents the summation for all over facets of the primary grid containing only $\{\pm 1, 0\}$, $\mathbf{T}$ is the transpose of $\mathbf{C}$ which represents summations of the facets of the dual grid, while $\mathbf{S}$ is the $\tilde{N}_V \times \tilde{N}_A$ matrix containing $\{\pm 1, 0\}$ and constructs the corresponding equation for each of the $N_V$ cells of the primary grid and $\tilde{\mathbf{S}}$ matrix for $\tilde{N}_V$ cells for the dual grid. Also the relations which transform quantities from the primary grid to the dual grid are needed for instance to transform electric and magnetic voltages into fluxes and vice versa (error terms are neglected for simplicity):

$$e_i = \int_{L_i} \mathbf{E} \cdot d\mathbf{s} = e \cdot L_i \qquad d_i = \int_{\tilde{A}_i} \varepsilon \mathbf{E} \cdot d\mathbf{A} = \varepsilon_{eff,i} \cdot e \cdot \tilde{A}_i \quad (5)$$

where $\varepsilon_{eff,i} = \frac{1}{\tilde{A}_i} \int_{\tilde{A}_i} \varepsilon dA$. Using the above relations, the discrete permittivity expression can be obtained:

$$\frac{d_i}{e_i} = \frac{\varepsilon_{eff,i} \, \tilde{A}_i}{L_i} \implies M_\varepsilon(i,i) = \frac{\varepsilon_{eff,i} \, \tilde{A}_i}{L_i} \qquad (6)$$

where $M_\varepsilon$ is a diagonal matrix which relates the voltages and fluxes. Assuming Cartesian grids for instance, the *x*-component of above expression reads:

$$M_\varepsilon(i,i) = \frac{\varepsilon_{eff,i} \, \Delta \tilde{y}_i \Delta \tilde{z}_i}{\Delta \tilde{x}_i} \qquad (7)$$

In a similar way, the relations between magnetic voltage along the dual grid and magnetic flux along the primary grid read:

$$\frac{h_i}{b_i} = \frac{\mu_{eff,i} \, \tilde{L}_i}{A_i} \implies M_\mu^{-1}(i,i) = \frac{\mu_{eff,i}^{-1} \, \tilde{L}_i}{A_i} \qquad (8)$$

where $M_\mu$ is the discrete permeability matrix which is also diagonal. Similarly, the *x*-component of the above expression for Cartesian grids reads:

$$M_\mu^{-1}(i,i) = \frac{\mu_{eff,i}^{-1} \, \Delta \tilde{x}_i}{\Delta \tilde{y}_i \Delta \tilde{z}_i} \qquad (9)$$

The equations Eq.(6-9) are called *discrete material relations* which are very useful for the implementation of the numerical scheme of FIT. The Maxwell's Grid Equations can be solved numerically using the leapfrog scheme which assumes staggered grid for the time-discretization with the application of central finite difference approximation for the time derivatives. Finally, assuming a Cartesian grid for the spatial variables and applying finite difference approximations, we obtain the well-known update equations of FDTD [5]. The spatial and time-domain discretized form of the Eq.(4) is therefore computationally equivalent to the FDTD approach. For a detailed description of the FIT scheme and the required spatial and time stability criteria of the algorithm, see Ref. [1-3].

## II.  Electron excitation and Thermal effects

To describe the ultrafast electron dynamics and the relaxation process for Au and Stainless steel, following irradiation with fs pulses, the conventional Two Temperature Model (TTM) is used. More specifically, the following set of equations is employed to investigate the spatio-temporal distribution of the produced thermalized electron ($T_e$) and lattice ($T_L$) temperatures of the asse

$$\begin{aligned}
C_e \frac{\partial T_e}{\partial t} &= \vec{\nabla} \cdot \left( k_e \vec{\nabla} T_e \right) - G_{eL}(T_e - T_L) + S(\vec{r},t) \\
C_L \frac{\partial T_L}{\partial t} &= \vec{\nabla} \cdot \left( k_L \vec{\nabla} T_L \right) + G_{eL}(T_e - T_L)
\end{aligned} \qquad (10)$$

where the subscripts *e* and *L* are associated with electrons and lattice, respectively, $k_e$ ($k_L$) is the thermal conductivity of the electrons (lattice) while $C_e$ and $C_L$ are the heat capacity of electrons and lattice, respectively, and $G_{eL}$ is the electron-phonon coupling factor. In particularly, $k_e \left( = k_{e0} \frac{B_e T_e}{A_e (T_e)^2 + B_e T_L} \right)$ (or lattice $k_L = 0.01 k_e$ ). In Eqs.10, $S(\vec{r},t)$ represents a term that is associated with the electromagnetic fingerprint of the material and the absorbed energy from the laser source. The values of the parameters used in the simulations are given in Table 1.

Simulations for the two materials demonstrate the significant impact of the thermophysical properties (and, more specifically, the electron-phonon coupling and electron heat conductivity). Figs. 1S-a,b show the spatial distribution of the lattice temperature for Au and Stainless steel (on

|  | Material | |
|---|---|---|
| **Parameter** | **Au** | **Steel** |
| $G_{eL}^{(m)}$ [Wm$^{-3}$K$^{-1}$] | Ab-Initio [6] | Ab-Initio [7] |
| $C_e^{(m)}$ [Jm$^{-3}$K$^{-1}$] | Ab-Initio [6] | Ab-Initio [7] |
| $C_L^{(m)}$ [×10⁶ Jm⁻³K⁻¹] | 2.48 [8] | 3.27 [7] |
| $k_{e0}^{(m)}$ [Wm⁻¹K⁻¹] | 318 [8] | 46.6 [7] |
| $T_{melt}$ [K] | 1337 [9] | 1811 [7] |
| $A_e$ [×10⁷ s⁻¹K⁻²] | 1.18 [9] | 0.98 [7] |
| $B_e$ [×10¹¹ s⁻¹K⁻¹] | 1.25 [9] | 2.8 [7] |

Table 1: Thermophysical properties of Au and Stainless steel.

the propagation plane), respectively, at $t$=20 ps following irradiation of a pattern that has been produced after five pulses ($NP$=5). Similarly, the spatial distribution of the electron temperature for Au and Stainless steel, respectively, are illustrated at time points shortly after which $T_e$ reaches the peak value (at $t$=1.3 ps and at $t$=1.7 ps, for Au and Stainless steel, respectively) (Figs. 1S-c,d). The (spatially) periodic distribution of $T_e$ (Figs. 1S-c,d) as a result of the periodic electromagnetic response (shown in Figs.1,2 in the main text) is transferred through electron

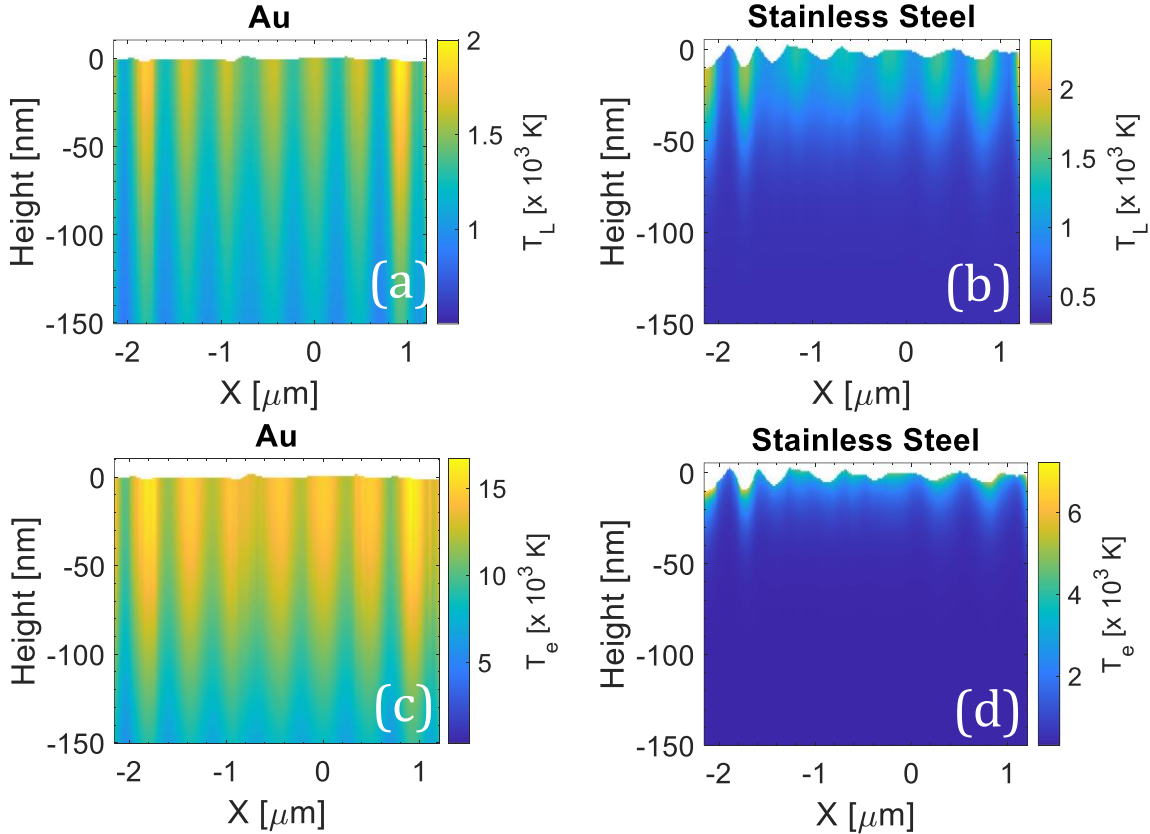

FIG.1S. Lattice temperature spatial profile at $t$=20 ps for (a) Au, (b) Stainless steel. Electron temperature spatial profile at $t$=1.3 ps for (c) Au and at $t$=1.7 ps for (d) Stainless Steel.

-phonon coupling to the lattice system (Figs. 1S-a,b). Furthermore, (as seen, also, in Fig.3 in the main manuscript), the electron diffusion for Au in larger than that for Stainless steel which indicates that the highly energetic electrons do diffuse to large depths that causes a shortage of electrons of high temperature that are capable to interact with the lattice system through electron-phonon scattering processes. The latter is expected to influence the spatial distribution of the lattice temperature field and the depth of the affected region (i.e. larger in Au). Furthermore, both the electron and lattice temperature gradients along the $X$-axis for Au are predicted to be smaller than those for Stainless steel (Figs. 1S) which indicates that the induced spatial periodicity does not survive to lead to periodic structures.

### III. Fluid Dynamics

One of the predominant effects that lead to surface modification is a phase transition the material undergoes upon relaxation. Despite other mechanisms can generate surface modification (such as ablation), in the current study, the focus is on laser conditions that lead to patterning due to a phase transition. More specifically, when the lattice temperature exceeds the melting point of the solid, $T_{melt}$ (Table 1) hydrothermal effects generate a mass displacement of the material that, in turn, leads to a patterned topography. The conventional approach that is used to describe the topography formation is through the analysis of the phase transition, fluid dynamics, evolution of the induced hydrothermal waves and finally, the resolidification process [10]. Assuming the molten material behaves as an incompressible fluid, the dynamics of the molten volume is described by the following Navier-Stokes equation (NSE)

$$\rho_0 \left( \frac{\partial \vec{u}}{\partial t} + \vec{u} \cdot \vec{\nabla} \vec{u} \right) = \vec{\nabla} \cdot \left( -P + \mu(\vec{\nabla}\vec{u}) + \mu(\vec{\nabla}\vec{u})^T \right) \qquad (11)$$

where $\rho_0$ and $\mu$ stand for the density and viscosity of molten material, while $P$ and $\vec{u}$ are the pressure and velocity of the fluid. In Eq.11, superscript $T$ denotes the transpose of the vector $\vec{\nabla}\vec{u}$ [10]. A more detailed description of the fluid dynamics module and numerical solution of NSE has been provided in various previous reports [10-12]. The values for the parameters (Eq.11) used in the simulations in this work are given in Table 2.

| Parameter | Material | |
|---|---|---|
| | Au | Steel |
| $\rho_0$ [kg m$^{-3}$] | 1.74×10$^4$ -1.44 ($T_L$-$T_{melt}$) [13] | 6900 [14] |
| $\mu$ [mPa s] | 4.240 | 16 [15] |

Table 2: Liquid phase properties of Au and Stainless steel.

## III. NUMERICAL SOLUTION

The above set of equations (Eqs.10-11) is characterized by an inherent complexity due to the high nonlinearity that is introduced by the electron temperature dependence of the thermophysical properties. Hence, an analytical solution is not feasible and therefore, a numerical approach is pursued. Numerical simulations have been performed using the finite difference method while the discretization of time and space has been chosen to satisfy the Neumann stability criterion. Furthermore, on the boundaries, von Neumann boundary conditions are applied while heat losses at the front and back surfaces of the material are assumed to be negligible. The initial conditions are $T_e(t=0) = T_L(t=0) = 300$ K and displacements are set to zero at $t=0$.

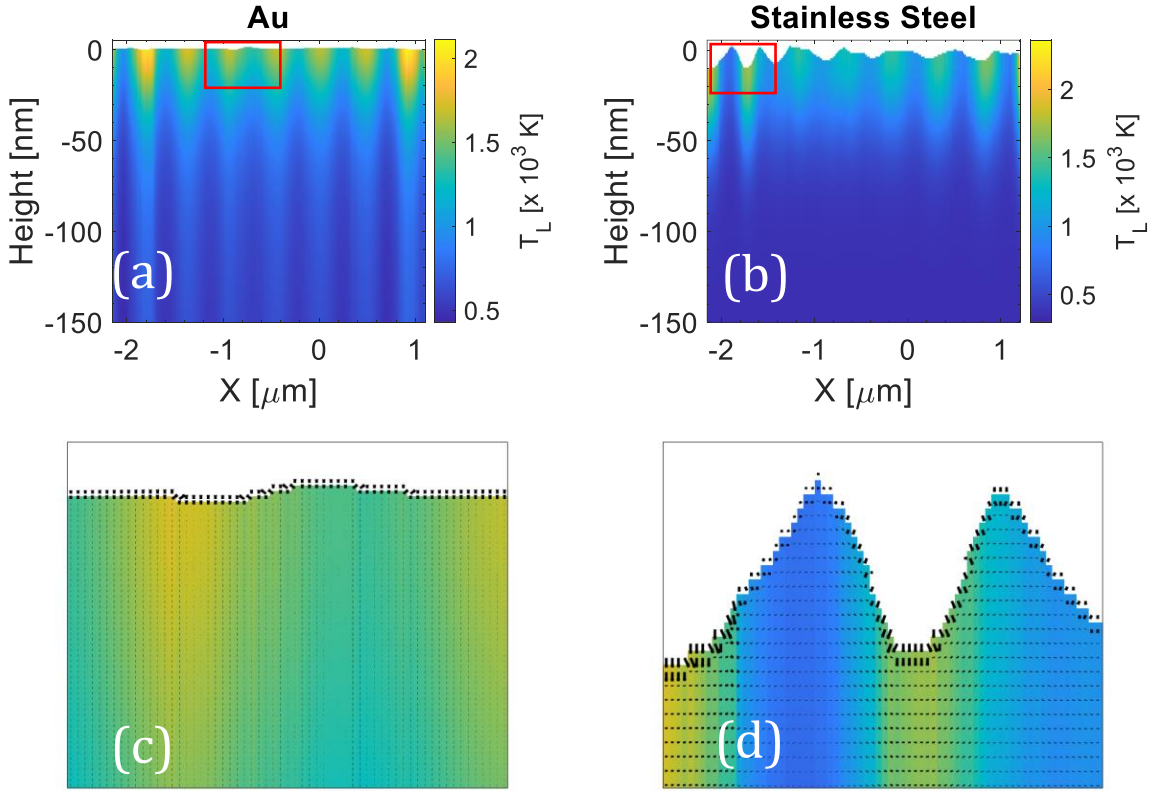

FIG.2S. Fluid movement at 20 ps for (a) Au and (b) Stainless steel. Figures (c) and (d) illustrate the fluid movement in the *red* rectangular region in (a) and (b), respectively (*NP*=5). *Black* arrows in (c), (d) indicate the movement direction and size of the arrows correspond to the magnitude of the fluid transport.

With respect to the investigation of the fluid dynamics, a precise estimate of the molten material behaviour requires a contribution from the surface tension related pressure, $P_\sigma$, which is influenced by the surface curvature and is expressed as $P_\sigma=K\sigma$, where $K$ is the free surface curvature and $\sigma$ surface tension. The calculation of the pressure associated to the surface tension requires the computation of the temporal evolution of the principal radii of surface curvature $R_1$ and $R_2$ that correspond to the convex and concave contribution, respectively [16]. The total curvature is computed by means of the expression $K=(1/R_1 +1/R_2)$ (see Ref. [10] for a detailed description of the simulation methodology). Pressure equilibrium on the material surface implies that the pressure $P$ in Eq.11 should outweigh the accumulative effect of $P_r+P_\sigma$. The thermocapillary boundary conditions imposed at the liquid free surface are the following [7, 17, 18]

$$\frac{\partial u}{\partial z} = -\sigma/\mu \frac{\partial T_L}{\partial x} \quad \text{and} \quad \frac{\partial v}{\partial z} = -\sigma/\mu \frac{\partial T_L}{\partial y} \tag{12}$$

where $(u,v,w)$ are the components of $\vec{u}$ in Cartesian coordinates. The Cartesian coordinate system indicated by $(x,y,z)$ is used to describe morphological changes compared to the initial $(x,y,z_S)$ for flat surfaces. The surface tension $\sigma$ (in Nm$^{-1}$) for the two materials are $1.93-1.73\times10^{-4}\,(T_L-T_{melt})$ K$^{-1}$ [19] and $1.15-1.6\times10^{-4}\,(T_L-T_{melt})$ K$^{-1}$ [20, 21] for Stainless steel and Au, respectively.

In Fig.2S, the lattice temperature profile is illustrated for Au and Stainless steel at $t=20$ ps, while the rectangular *red* box (Fig.2S-a,b) defines the region of interest where the fluid direction is illustrated in Fig.2S-c,d (*black* arrows indicate the fluid movement direction). Similar behaviour is derived in the rest of the region (results are not shown), however, due to the small size of the arrows, the arrow direction is not visible. It is noted that in order to simulate the fluid movement, we describe both the solid and liquid phase with a molten volume of substantially different viscosity (for more details, see Ref. [10]).